Research article

**Open Access**

# Context-dependent selection of visuomotor maps
## Emilio Salinas*

Address: Department of Neurobiology and Anatomy, Wake Forest University School of Medicine, Winston-Salem, NC, 27157-1010, USA

Email: Emilio Salinas* - esalinas@wfubmc.edu

* Corresponding author

## Abstract

**Background:** Behavior results from the integration of ongoing sensory signals and contextual information in various forms, such as past experience, expectations, current goals, etc. Thus, the response to a specific stimulus, say the ringing of a doorbell, varies depending on whether you are at home or in someone else's house. What is the neural basis of this flexibility? What mechanism is capable of selecting, in a context-dependent way an adequate response to a given stimulus? One possibility is based on a nonlinear neural representation in which context information regulates the gain of stimulus-evoked responses. Here I explore the properties of this mechanism.

**Results:** By means of three hypothetical visuomotor tasks, I study a class of neural network models in which any one of several possible stimulus-response maps or rules can be selected according to context. The underlying mechanism based on gain modulation has three key features: (1) modulating the sensory responses is equivalent to switching on or off different subpopulations of neurons, (2) context does not need to be represented continuously, although this is advantageous for generalization, and (3) context-dependent selection is independent of the discriminability of the stimuli. In all cases, the contextual cues can quickly turn on or off a sensory-motor map, effectively changing the functional connectivity between inputs and outputs in the networks.

**Conclusions:** The modulation of sensory-triggered activity by proprioceptive signals such as eye or head position is regarded as a general mechanism for performing coordinate transformations in vision. The present results generalize this mechanism to situations where the modulatory quantity and the input-output relationships that it selects are arbitrary. The model predicts that sensory responses that are nonlinearly modulated by arbitrary context signals should be found in behavioral situations that involve choosing or switching between multiple sensory-motor maps. Because any relevant circumstantial information can be part of the context, this mechanism may partly explain the complex and rich behavioral repertoire of higher organisms.

## Background

The concept of a direct, one-to-one association between a sensory stimulus and a motor response has been strongly influential in neuroscience [1]. Such associations may be quite complex; for instance, monkeys can learn visuomotor mappings based on arbitrary rules [2-4]. But from a mechanistic point of view, it is their flexibility which is remarkable. Humans and other mammals react to a given stimulus in drastically different ways depending on the context [1,5-7]. What is the neural basis for this? How do current goals, recent events, and other environmental circumstances gate or route immediate sensory signals to generate an adequate action?

Gain control is a common mechanism by which neurons integrate information from multiple modalities or sources [8,9]. Gain-modulated neurons typically have a sensory receptive field, but in addition, their overall excitability depends on some other modulatory parameter. A classic example are the neurons in parietal area 7a, whose activity can be described by the product of a gain factor, which is a function of the gaze angle, and the response profile of the visual receptive field [10,11]. That is, gaze direction determines the amplitude of their stimulus-dependent responses. According to theoretical studies, gain-modulated responses are useful for performing a class of mathematical operations known as coordinate transformations [12-16]. For example, by combining multiple eye-centered inputs that are gain modulated by gaze direction, a downstream neuron can generate a response that depends on the location of a stimulus relative to the body [12-14]. Experimental studies have reported gain changes due to a wide range of proprioceptive signals, such as gaze direction [10,11,17], eye and head velocity [18] and arm position [19,20]. Modulations relevant to attention-centered [21-23] or object-centered representations [24,25] have also been documented.

Interestingly, all of these examples deal with the same problem – spatial localization – but the computations that can be effectively carried out through gain-modulated responses are much more general [13,16,26]. In particular, here I show that modulating the activity of a population of neurons is equivalent to turning on and off different subsets of neurons. Thus, the modulation can be thought of as a switch that can activate one of many possible sensory networks, each instantiating a different sensory-motor map. Crucially, the modulatory signal itself does not have to provide any spatial information; it can be a recent instruction or some other aspect of the current behavioral context. Examples of choices between multiple sensory-motor maps determined in a context-dependent manner include speaking in one language or another, and the ability of musicians to interpret a musical score depending on the clef and key signature at the beginning of each stave. But the same principles also apply in more simplified settings, such as behavioral tasks where a given stimulus is arbitrarily associated with two or more motor responses, depending on a separate instruction [4,27-29]. For instance, the shape of a fixation point may be used to indicate whether the correct movement should be a saccade toward a spot of light or an antisaccade away from it [30]. What all of these cases have in common is a functional reconnection between visual and motor networks that must occur very quickly and without explicit spatial guidance from the context information.

Using theoretical and computer-simulation methods, I show that this type of functional switching can be achieved through contextual modulation regardless of how the context is encoded – whether continuously or discontinuously – and independently of the discriminability of the stimuli. The results are presented using neural network models of hypothetical behavioral tasks similar to those used in experiments with awake monkeys. A report with a different example was published previously [31].

## Results

All model networks discussed below have the same general, two-layer architecture [14-16]. A first layer of gain-modulated (GM) neurons drives a second layer of output or motor neurons through a set of feedforward connections, with each GM unit projecting to all output units. In each trial of a task, the GM neurons are activated by the sensory and context signals, and a motor response is generated by the output neurons (see Methods). Each model proceeds in three steps. First, the GM and the desired output responses are specified according to the task. Then, synaptic weights are found that, across all stimulus and context combinations, make the driven output responses as close as possible to the desired ones. Finally, the network is tested in multiple trials in which the GM neurons drive the output units. Model performance is measured by comparing the resulting, driven pattern of motor activity in each trial with the desired, pre-specified one. The first task, with only two contexts, serves to illustrate the analogy between gain modulation and a switch.

### Switching between saccades and antisaccades

In the antisaccade task, a stimulus appears briefly at position $x$ along the horizontal and the subject responds by making an eye movement (Fig. 1). There are two possible contexts or conditions. In the first one, the movement should be to the location where the stimulus appeared, $x$; in the second one, the movement should be to the mirror-symmetric point, $-x$. Both condition and stimulus location vary across trials. The color of the fixation spot (or any other arbitrary cue) may serve to indicate which condition applies in each trial [30].

Examples of model GM responses chosen for this task are shown in Fig. 2. These neurons simply respond to visual stimuli presented at different locations; however, they are also sensitive to the context. Each graph shows the mean firing rate of one unit as a function of $x$, with one curve for each of the conditions (red and green traces). These tuning curves are bell-shaped because Gaussian functions were used to define them (see Methods). Because context affects the gain of the responses, for any given cell, the two curves differ only in their amplitudes. The context that produces the highest gain is the preferred one. The maximum and minimum gains for each neuron are model parameters that can be between 0 and 1. The four GM units in Fig. 2 illustrate various degrees of modulation. The case of full modulation (maximum gain = 1, minimum gain = 0) depicted in Fig. 2a corresponds to a neuron that is switched on and off by context: in its preferred condition it is highly active, whereas in its non-preferred condition it is fully suppressed.

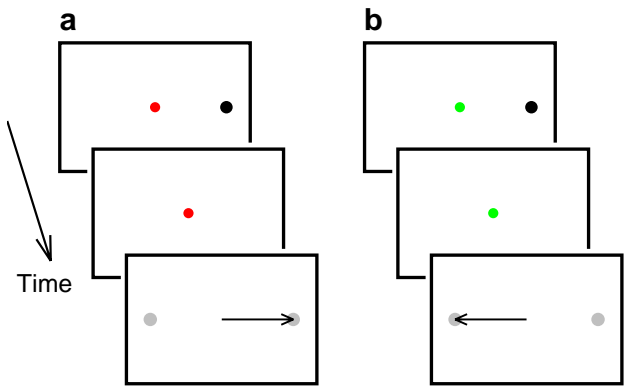

**Figure 1**
**Antisaccade task.** In each trial, a stimulus (black dot) is presented at a distance *x* from the fixation point (colored dot); the stimulus disappears; two targets appear (gray dots) and the subject responds by making an eye movement (arrow) to one of them. The color of the fixation spot indicates whether the movement should be a saccade or an antisaccade. **a**: In context 1 the fixation spot is red and the movement is to the target at *x*. **b**: In context 2 the fixation spot is green and the movement is to the opposite target, at -*x*. In the model, *x* is between -15 and +15, with distance in arbitrary units.

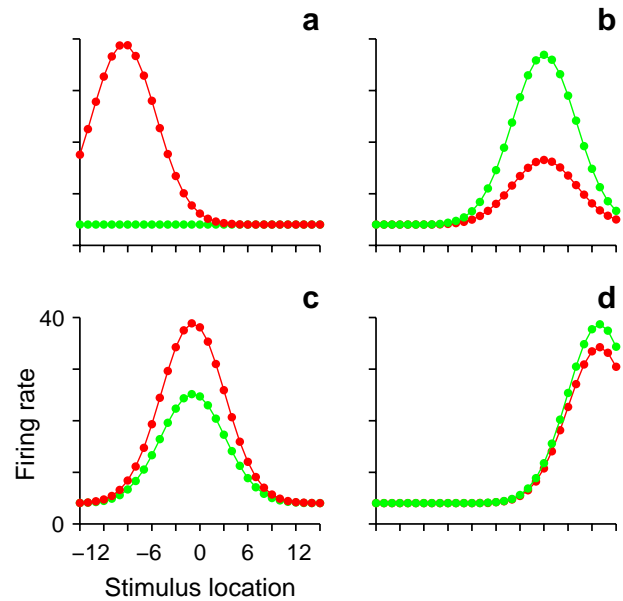

**Figure 2**
**Responses of GM neurons in the antisaccade task.** Each graph plots the mean firing rate of a model neuron as a function of stimulus location. Red and green traces correspond to sensory responses evoked during contexts 1 and 2, respectively. The gain in the preferred condition is 1. **a**: A unit that prefers context 1 and is 100% suppressed in the non-preferred condition; its minimum gain is $\gamma = 0$. **b**: A unit that prefers context 2 and is 62% suppressed in the non-preferred condition; its minimum gain is $\gamma = 0.38$. **c**: Another unit that prefers context 1; its minimum gain is $\gamma = 0.61$. **d**: Another unit that prefers context 2; its minimum gain is $\gamma = 0.87$. Firing rates are in spikes/s. Model responses are based on Equations 1 and 3.

First consider what happens if the first layer of a model network is composed of two populations of such switching neurons. One population is active in context 1 and the other in context 2. This is illustrated in Fig. 3a. The rectangle encloses the responses of all model neurons (60 GM and 25 output units) in a single trial of the antisaccade task. The firing rates of the GM neurons are in color. The two populations (red and green) have opposite context preferences but identical sets of sensory tuning functions. The black dots are the responses of the driven output neurons. Their center of mass (Equation 19), which in this case is the same as the location of the peak, is interpreted as the target location for an impending saccade. The network performs accurately in the four trials shown in the column, since the encoded movement location is equal to *x* for saccades (context 1) and to -*x* for antisaccades (context 2). It is easy to see why such a network can implement two entirely independent sensory-motor maps: each population has its own set of synaptic connections driving the downstream motor neurons, and the maps are kept separate because the two populations are never active at the same time.

Figure 3d shows the corresponding matrix of synaptic connections. To interpret this figure, notice that GM units 1–30 are the ones that prefer context 1 (red dots in Fig. 3a), whereas units 31–60 prefer context 2 (green dots in Fig. 3a). Preferred stimulus locations are arranged in increasing order for both populations. Units 1–30 generate direct saccades, so their connections are aligned with the motor neurons; that is, GM neuron 1 excites output neuron 1 most strongly, GM neuron 2 excites output neuron 2 most strongly, etc. Thus, in context 1, stimuli to the right generate movements to the right. In contrast, the GM units that generate antisaccades are connected in the reverse order: GM neuron 31 excites output neuron 25 most strongly, GM neuron 32 excites output neuron 24 most strongly, and so on. Thus, in context 2, stimuli to the right result in movements to the left.

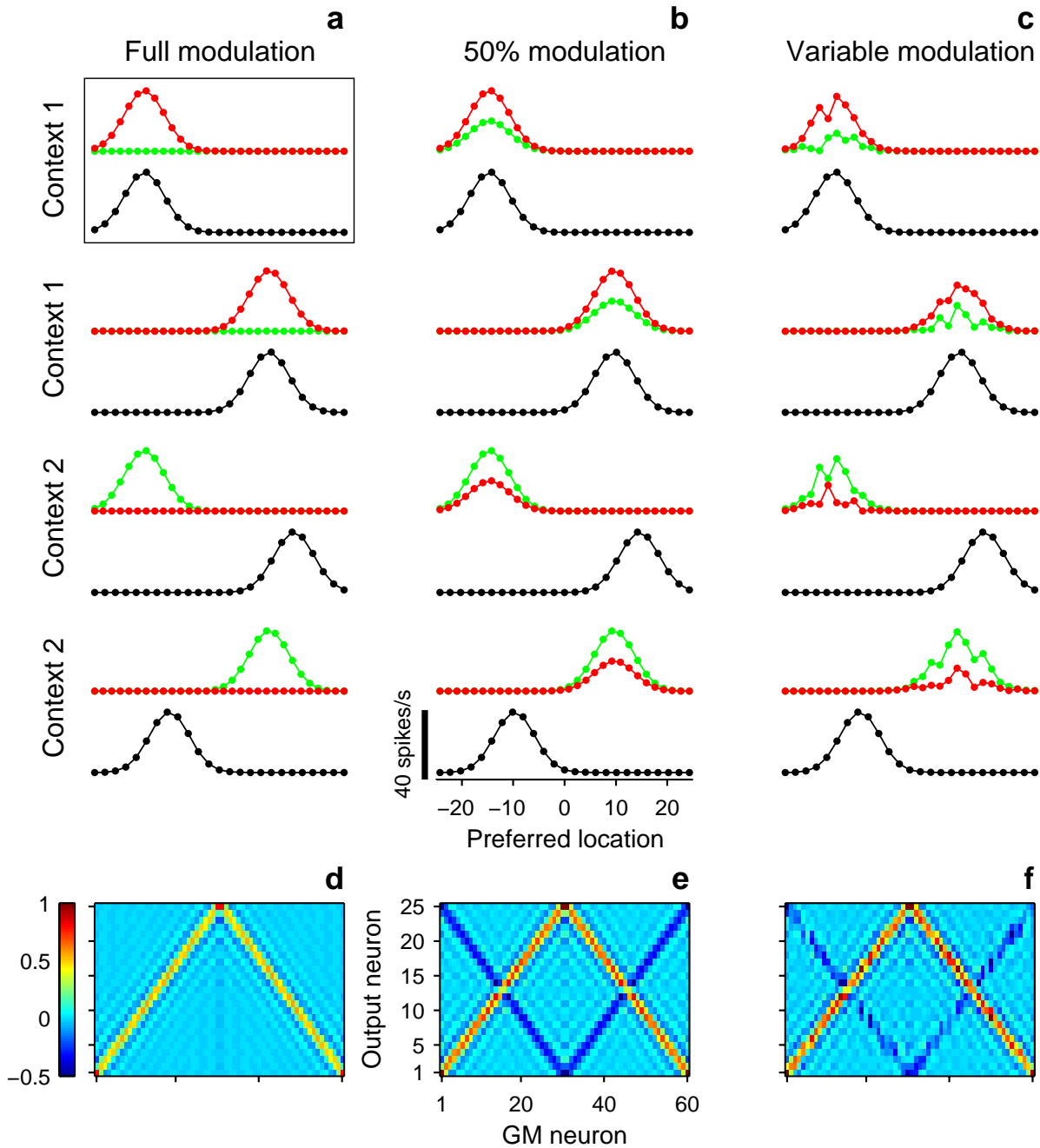

**Figure 3**
**Network performance in the antisaccade task. a**: Firing rates of all model cells when the GM units are fully modulated ($\gamma$ = 0). The box marks a single trial. Colored traces are the firing rates of the 60 GM neurons in the network; 30 of them (red) prefer the direct saccade condition and 30 (green) prefer the antisaccade condition. Black dots are the 25 motor responses driven by the GM neurons. For GM responses, x-axis is preferred stimulus location; for output responses, x-axis is preferred movement location. Context in each of the four trials is indicated on the left. Trials with $x$ = -15 and $x$ = 10 alternate. The profile of output activity always peaks at the correct location. **b**: As in **a**, but when all GM neurons are partially modulated by the same amount ($\gamma$ = 0.5). **c**: As in **a**, but when the maximum and minimum gains of the GM units are chosen randomly from uniform distributions. **d-f**: Connection matrices for the networks in the respective columns. Each point shows the synaptic weight, coded by color, from one GM neuron to one output neuron. GM units 1–30 (red points in upper panels) prefer context 1, whereas GM units 31–60 (green points in upper panels) prefer context 2. No noise was included in the simulations ($\alpha$ = 0).

The model correctly produces saccades in context 1 and antisaccades in context 2. Furthermore, this scheme for switching sensory-motor maps as a function of context would also work for any two maps driven by the two populations. This model switches maps successfully because the GM neurons are themselves switched on and off by context, so this case is trivial. However, the main result in this section is that a network of partially modulated GM neurons has exactly the same functionality. The more rigorous statement is this: for a discrete number of contexts and everything else being equal, a network of partially modulated neurons can generate the same mean downstream responses as a network of switching neurons. Figure 3 illustrates this equivalence: identical output activity profiles are generated when all GM neurons are fully suppressed in their non-preferred context (Fig. 3a), when all are partially modulated by the same amount (Fig. 3b), and when the modulation varies randomly across cells (Fig. 3c). These three cases require different sets of synaptic connections between GM and output layers, but this is simply because the GM responses vary across cases. In particular, note the dark blue diagonal bands in Figs. 3e,3f, compared to Fig. 3d. They correspond to negative weights needed to subtract out activity that is irrelevant to a particular context. For instance, in the direct saccade trials of Fig. 3b, the responses of the antisaccade-preferring neurons should be cancelled, and viceversa. The new negative weights combined with larger positive weights achieve this.

The key point is that, under relatively mild conditions, partial and full modulation lead to the exact same repertoire of switchable sensory-motor maps (the difference lies in their accuracy, as discussed below). The formal proof is presented in Appendix A. This result is interesting because it provides an intuitive interpretation of gain modulated activity: modulations that may seem small at the single-unit level may produce drastically different output responses due to their collective effects, the result being as if different sensory populations had been turned on and off.

### Partial versus maximum gain modulation

The equivalence between networks of neurons that switch across contexts and networks with partial modulation is subject to an important condition and a qualification.

The key condition is that the modulation factors that determine the gain of all the neurons with similar stimulus selectivities must be linearly independent across contexts (Appendix A). In practice, one way to achieve this is to include all relevant combinations of sensory and contextual preferences. For instance, if there are two neurons that respond maximally when $x = 5$, the condition is fulfilled for that pair if one neuron prefers context 1 and the other context 2. As long as this independence constraint is satisfied, there is great flexibility in the actual amount of modulation; it does not need to be 100%, as with a full switch.

The qualification, however, is also critical, because a network of partially modulated GM neurons is not exactly the same as one composed of switching neurons: in most functionally relevant cases, partially modulated neurons are effectively noisier. In general, variability plays an important role in the performance of these networks. No fluctuations were included in the simulations of Fig. 3, so performance was virtually perfect. But the magnitude of the error between correct and encoded movement directions increases depending on the amount of noise that is added to the GM responses, and as the difference between the minimum and maximum gains diminishes, the impact of noise typically goes up. This is shown analytically in Appendix C and is illustrated in Fig. 4.

Two measures of noise sensitivity are plotted in Fig. 4. The first one is the standard deviation of a single output response across trials with identical stimulus and context. This number, $\sigma_R$, quantifies the variability of single neurons. Figure 4a plots $\sigma_R$ as a function of $\gamma$, which is the minimum gain of the GM neurons (the maximum is 1). When $\gamma = 0$, the GM neurons are fully suppressed in their non-preferred context; when $\gamma = 1$, the GM responses are identical in both contexts. The three curves are for three levels of noise. Their order shows that, as expected, higher noise in the input layer always produces higher variability in the output. For each data point, the synaptic weights were set so that the average firing rates of the output neurons, as functions of stimulus location and context, were always the same (Appendix B). Thus, for all $\gamma$ values, the average profile of motor responses for $x = -15$ and $x = 10$ looked exactly like those in Fig. 3. The monotonically increasing curves in Fig. 4a indicate that the variability of the output rates goes up with $\gamma$, as predicted theoretically (Appendix C).

The second measure of noise sensitivity is $\sigma_{CM}$, which estimates the error between the desired movement location and the center of mass of the output population, which is considered the encoded movement location (Equations 19, 20). Thus, $\sigma_{CM}$ quantifies the variability of the network. Figure 4b shows that $\sigma_{CM}$ also increases with $\gamma$, reaching a saturation level. This error saturates because, in contrast to the individual neuron responses, the encoded movement location is restricted to a limited range of values, so its variance cannot grow above a certain limit.

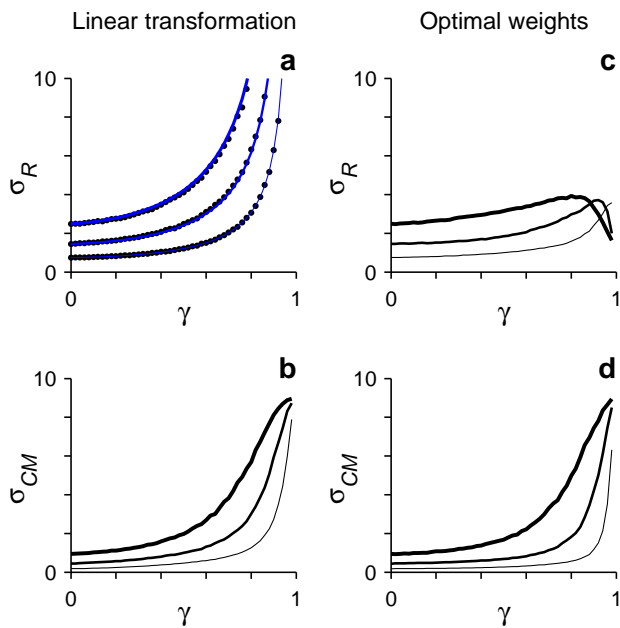

**Figure 4**
**Sensitivity to noise as a function of modulation strength in the antisaccade task.** All results are for networks of 60 GM and 30 output neurons. The x-axes indicate $\gamma$, which is the minimum gain of the GM neurons; the maximum is always 1. In all panels, the three curves are for three levels of noise: $\alpha$ = 0.04 (thin lines), $\alpha$ = 0.36 (medium lines), or $\alpha$ = 2.25 (thick lines). **a**: Standard deviation of single output firing rates, averaged over stimulus locations and contexts, as a function of $\gamma$. Data points are from simulations; continuous lines are analytic results from Equation 39, with $a$ = 1.42. For each data point, the average output responses, as functions of $x$ and $y$, were the same. To achieve this, the synaptic weights for $\gamma > 0$ were obtained by a linear transformation of the weights for $\gamma = 0$ (Appendices B, C). **b**: Error between correct and encoded movement locations as a function of $\gamma$. Results are from the same simulations as in **a**. **c, d**: As in **a, b**, respectively, but for simulations in which the synaptic weights were computed using the standard, optimal algorithm (see Methods). Note that $\sigma_{CM}$ always increases with $\gamma$.

Figures 4c and 4d show the same measures of variability but when the synaptic weights are computed using the standard, optimal algorithm (see Methods). For each value of $\gamma$, the optimal algorithm considers both the mean and the variance of the output responses [32,33], striking a balance between them that, overall, minimizes the average squared difference between the driven and the desired output rates (Equation 11). Therefore, in Figs. 4c,4d, the mean output responses are not quite the same for all data points; in particular, for $x$ = -15 and $x$ = 10 there are small differences compared to the curves in Fig. 3 (data not shown). This method markedly reduces the variability of the individual output neurons relative to the case where only the mean values are considered. It also produces a modest decrease in $\sigma_{CM}$ (compare Figs. 4b and 4d). However, it does not change the main effect: the error in the encoded location still grows monotonically with $\gamma$.

Note that, as explained in Appendix C, $\gamma > 0$ does not always produce higher variance in the output, compared to $\gamma = 0$. For instance, if the sensory-motor maps in the two contexts are the same, the optimal strategy is to activate both populations of GM neurons simultaneously, i.e., to use $\gamma = 1$. This is simply because the average of two noisy responses with equal means is better than either of them. In general, however, switching is relevant precisely when the sensory-motor maps are different, as in Figs. 3 and 4, in which case weaker modulation (higher $\gamma$) results in higher output variability.

In conclusion, as the modulation becomes weaker, the performance of the network typically becomes less accurate, even though the average output responses may be close or identical to those obtained with maximum modulation. In Fig. 4, this becomes more of a problem when the minimum gain $\gamma$ is above 0.6 or so, at which point $\sigma_{CM}$ is about twice that observed with full modulation. These results were obtained using the same $\gamma$ for all GM neurons, but almost identical curves were produced when $\gamma$ varied randomly across cells and the results were plotted against its average value.

*Continuous vs discontinuous context representations*
The possible contexts encountered by an organism could be numerous and diverse, so it is not clear how the brain might encode them. There are at least two distinct ways: as separate, discrete states, or as points along a smooth, continuous space. What would be the difference in terms of the functionality of the remapping networks studied here? This is investigated next, using a generalization of the antisaccade task referred to as the scaling task.

The scaling task is very much like the antisaccade task, except with more contexts. The subject's response should be an eye movement toward a location determined by the position of the stimulus, $x$, and a scale factor, $y$; the movement should be toward the point $xy$. When $y = 1$, the movement is simply a saccade toward $x$; when $y = -1$, the movement is an antisaccade toward $-x$; when $y = 0.5$, the movement should be to a point halfway between fixation and the location of the stimulus, and so on. To begin with,

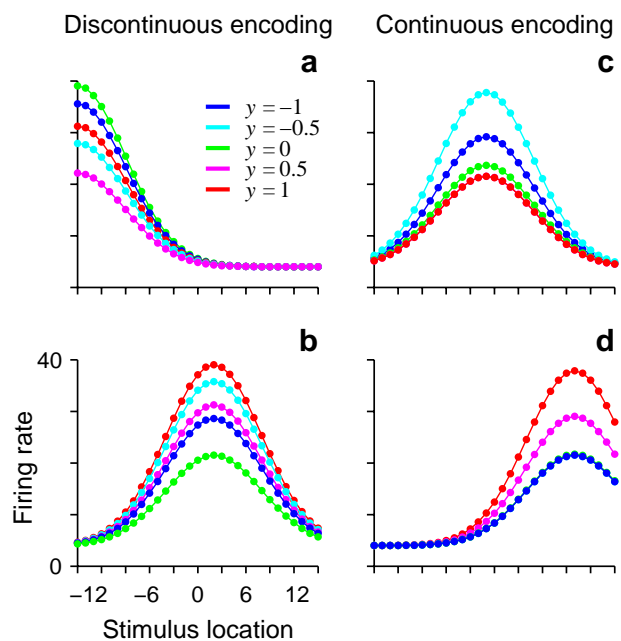

**Figure 5**
**Responses of four model GM neurons in the scaling task.** Same format as in Fig. 2, except that there are five possible contexts, corresponding to $y$ = -1, -0.5, 0, 0.5 and 1. **a**: Tuning curves for a model neuron that responds maximally to stimuli at $x$ = -15 and prefers the green condition ($y$ = 0). The order of effectiveness for the five scales was set randomly, so context is encoded discontinuously. **b**: As in **a**, but for another neuron that prefers $x$ = 2 and $y$ = 1. **c**: Tuning curves for a model neuron that encodes context in a smooth, continuous way. The unit responds maximally to stimuli at $x$ = -1 and prefers the cyan condition ($y$ = -0.5). The gain of the cell decreases progressively as $y$ deviates from the preferred scale – note the order of the colors. **d**: As in **c**, but for a neuron that prefers $x$ = 9 and $y$ = 1. All units have a maximum gain of 1 and a minimum gain near 0.5. Model responses were based on Equations 1, 3, 5 and 6.

five possible conditions are considered, corresponding to scales of -1, -0.5, 0, 0.5 and 1.

Figure 5 shows the responses of four GM units in this task plotted as functions of the position of the stimulus along the horizontal. A family of five curves, one per condition or scale factor, is drawn for each unit. As in the previous task, the shape of these curves is constant across conditions, because of the multiplicative interaction between stimulus- and context-dependent factors. The neurons in Figs. 5a,5b encode the context in a discontinuous way, because the order in which they prefer the five scales was set randomly (see Methods). Thus, for each unit, the order of the colors in Figs. 5a,5b is random. In contrast, the neurons in Figs. 5c,5d encode context smoothly; their response amplitudes decrease progressively as the current scale $y$ differs from each cell's preferred scale. All units in the figure have approximately the same minimum gain, $\gamma \approx 0.5$.

Differences between these two coding strategies can be observed in Fig. 6. This figure shows the performance of two versions of the network model, each with 900 GM cells, in four trials of the scaling task. In the first version, illustrated in Figs. 6a-6d, context is encoded discontinuously, as in Figs. 5a,5b. The GM firing rates are color-coded, ordered according to their preferred stimulus locations (x-axis) and preferred scales (y-axis). In each trial, the GM rates form a band of activity centered on the location of the stimulus. The most intense responses are somewhat clustered, although high firing rates are scattered throughout the band. The band occurs because the responses vary smoothly as functions of stimulus location, and the scatter in the y-direction is due to the random order in which each neuron prefers the contexts; such scatter would be present even without noise. The output neurons have profiles of activity (black traces) with the highest peak located near the intended movement target. The small wiggles and secondary bumps are due to noise. The performance of the network is accurate, however: the encoded movement is close to the intended one for all combinations of stimulus location and scale factor (Figs. 6a-6d, red vs black vertical lines). The second version of the model, illustrated in Figs. 6e-6h, is almost identical to the first, except that context is encoded continuously, as in Figs. 5c,5d. Now the the activation pattern that emerges is clearly localized, centered on the current stimulus and context values. Performance is similar for the two networks, both having $\sigma_{CM} \approx 0.6$.

Figures 7a,7b evaluate the performance of these two models across a wider range of parameters. The graphs show $\sigma_{CM}$ as a function of the number of GM neurons for three levels of noise. In all cases, the error decreases approximately as $1/\sqrt{N}$ – a sign that noise is what limits the accuracy of the system. This is consistent with the virtually perfect performance obtained with zero noise. With the five selected contexts, results are almost identical for the continuous and discontinuous cases.

### *Robustness and generalization*
There are two aspects of these networks that could vary depending on how context is encoded. The first is their robustness. In addition to random variations in the GM responses, there could be fluctuations in other elements of the circuits, in particular, the synaptic connections. Thus, a key question is how finely-tuned these connections need to be in order to obtain accurate per-formance. The answer: not very much. The networks tolerate considerable alterations in synaptic connectivity. This is illustrated in Figs. 7c,7d, which show $\sigma_{CM}$ as a function of the number of GM neurons in networks in which the connections were corrupted. For these plots, the connections were first set to their optimal values, as in the standard simulations, but then 25% of them, chosen randomly, were set to zero. To generate the same range of output firing rates, all remaining connections were divided by 0.75, but no further adjustments were made. Performance was then tested. Compared to the results with unaltered weights (Figs. 7a,7b), performance is evidently worse, but the disruption is not catastrophic; in particular, the error still goes down with network size. The increase in error is most evident when the noise is relatively low. Random weight deletion was used for these simulations because it is a rather extreme form of weight corruption, but other manipulations generated similar results.

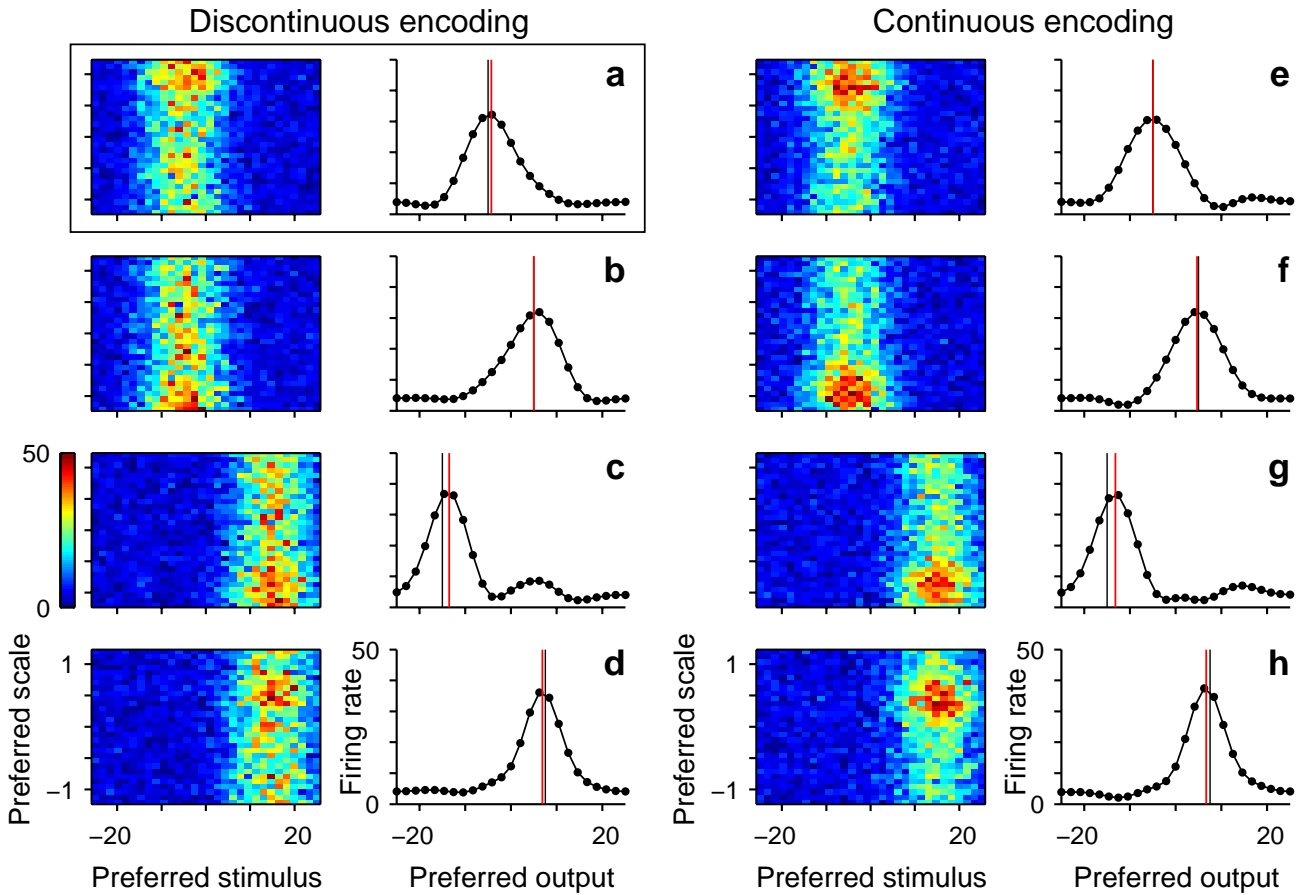

**Figure 6**
**Network performance in the scaling task.** Results are from two networks, one that encodes context discontinuously (first two columns) and another that encodes it continuously (third and fourth columns). **a**: The box encloses all model responses in a single trial with $x = -5$ and $y = 1$. The color plot shows all 900 GM responses, color coded. Neurons are arranged by preferred stimulus location along the x-axis and by preferred context along the y-axis. Black traces are the firing rates of the 25 driven output neurons. The black line indicates intended target location ($xy = -5$); the red line indicates encoded target location (center of mass). Their difference (error) is -0.73. **b**: A trial with $x = -5$, $y = -1$ and error = 0.02. **c**: A trial with $x = 15$, $y = -1$ and error = -1.54. **d**: A trial with $x = 15$, $y = 0.5$ and error = 0.63. **e-h**: Same combinations of stimulus and context as in **a-d**, but using a smooth representation for context. Errors are -0.07, 0.34, -1.84, and 0.8, respectively. The variance of each GM rate is equal to its mean ($\alpha = 1$).

The second important issue about these networks is their capacity to generalize. So far, the models have been tested with the same stimuli and contexts used to set the connections, but what happens when new stimuli or contexts are presented? This is where partial modulation and a smooth organization of response properties make a difference. First consider the model in which scale is encoded discontinuously. Its performance in generalization is shown in Fig. 7e. For this graph, only 8 stimulus locations, in combination with the 5 possible scales, were used to calculate the synaptic weights. That is, only 8 evenly-distributed values of *x* were used in Equation 3, giving a total of 40 combinations of stimulus and context. However, the network was tested with all 151 combinations of 31 stimulus locations (between -15 and +15) and 5 scales. Accuracy is practically the same as in the original simulations (Fig. 7a), where the 31 stimuli and 5 scales were used both for setting the connections and evaluating performance. The same scales had to be used in both cases because, given the discontinuous encoding, the gain factors for other scales could not be interpolated or inferred.

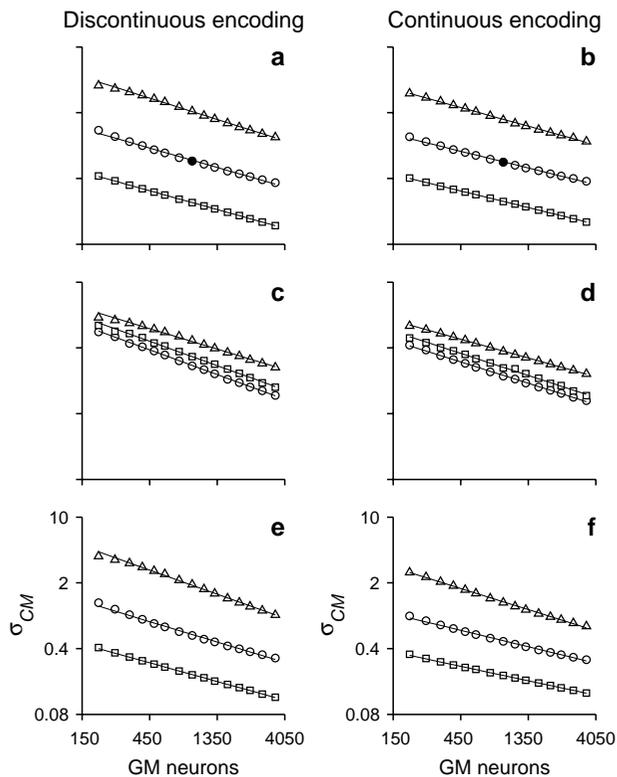

**Figure 7**
**Robustness and generalization in the scaling task.** Left and right columns are for networks in which scale is encoded discontinuously and continuously, respectively. Each panel shows results for three noise levels: $\alpha$ = 0.09 (squares), $\alpha$ = 1 (circles) and $\alpha$ = 9 (triangles). **a**: Error in encoded movement location as a function of the number of GM neurons. Each point represents an average over stimulus locations, scales and trials; 31 stimulus locations and 5 scales were used both to set the connections and test the networks. The filled symbol indicates the network in Figs. 6a-d. **b**: As in **a** but for networks in which scale is encoded continuously. The filled symbol indicates the network in Figs. 6e-h. **c, d**: $\sigma_{CM}$ vs network size in networks with corrupted synaptic weights. These simulations proceed as in **a, b**, except that performance is tested after deleting 25% of the synaptic connections, chosen randomly. **e**: $\sigma_{CM}$ vs network size when only 8 stimulus locations (combined with the 5 scales) are used to set the connections and the network is tested with all combinations of 31 stimulus locations and 5 scales. **f**: $\sigma_{CM}$ vs network size when combinations of only 8 stimulus locations and 8 scales are used to set the connections and performance is evaluated with all combinations of 31 stimulus locations and 31 scales. Continuous lines are linear fits to the data points above 250 units. Note logarithmic axes.

In contrast, in the continuous case, generalization can be tested in both the sensory and modulatory dimensions; the GM responses can be obtained for any combination of stimulus location and scale, because both quantities are defined analytically (Equations 3 and 6). Results are shown in Fig. 7f. For this graph, 8 stimulus locations and 8 scale factors were used to set the connections. The network was then tested on 31 stimulus locations and 31 scales within their respective ranges. Performance is slightly better than in the standard condition in which identical combinations of 31 stimulus locations and 5 scales were used throughout (Fig. 7b).

In summary, this task requires somewhat more complex GM neurons than the antisaccade task, because there are more contexts. In the discontinuous case, the basic intuition for why the model works is the same as in the previous task: with the proviso that they are effectively noisier, partially modulated neurons are equivalent to switching neurons, which can trivially establish independent sensory-motor maps. However, the key advantage of a continuous neural representation of context over a discontinuous one is that it allows a network to perform accurately on combinations of stimulus and context that have not been explicitly encountered before. By its very definition, such continuous encoding requires partial modulation. Therefore, although partial modulation is typically detrimental for switching between discrete contexts (relative to full switching), it is highly advantageous when context is parameterized by a continuous variable, because it serves to generalize.

### *Remapping based on ambiguous stimuli*
In the scaling task, all stimuli and contexts are unambiguous, but in many real-life situations and experimental paradigms, motor actions are preceded by perceptual processes that involve the interpretation or analysis of sensory information. That is, specific actions (e.g., pressing a left or right button) are often based on ambiguous information (e.g., whether on average a group of flickering dots moved to the left or to the right). In theory, switching between maps should be independent of the perceptual component of a task (Appendix A).

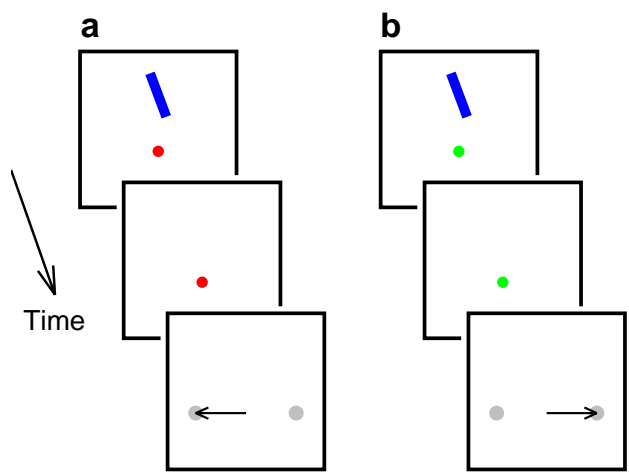

**Figure 8**
**Orientation discrimination task.** In each trial, a bar oriented at an angle x is presented while the subject fixates; the stimulus disappears; two targets appear (gray dots), and the subject indicates whether the bar was tilted to the left or to the right by making an eye movement (horizontal arrow). Vertical bars correspond to x = 0°. **a**: The bar is tilted to the left (x < 0). With a red fixation spot (y = 1), responses to left- and right-tilted bars should be to the left and right targets, respectively. The correct response is thus to the left. **b**: With a green fixation spot (y = 2), left- and right-tilted bars correspond to right and left targets, respectively. The correct response is now to the right. A no-go condition (y = 3; not shown) is included in the simulations in addition to the two go conditions. Orientation is in degrees, with x between -8° and +8°.

To investigate this, consider the orientation discrimination task illustrated in Fig. 8. In each trial, a bar is presented and the subject must determine whether it is tilted to the left or to the right. The judgement is indicated by making an eye movement either to a left or a right target. Discrimination difficulty varies depending on orientation angle $x$. The task is most difficult when $x$ is near 0° and the bar is nearly vertical, but it becomes easier as $x$ approaches ± 45°. This is also a remapping task because the association between bar orientation and correct target is not unique: the color of the fixation spot determines whether left and right targets correspond to bars tilted to the left ($x < 0$) and to the right ($x > 0$), respectively, or viceversa. There is also a no-go condition, which gives a total of three.

The GM cells in this case are tuned to stimulus orientation. The response curves are not shown, but have a single peak, as in Figs. 5a,5b – the difference is that the sensory variable is orientation, which varies from -90° to +90°, and that there are only three conditions, three values of $y$ (see Methods). The order in which each GM cell prefers the three contexts is set randomly, so context is encoded discontinuously.

The responses of the model output units are shown in Figs. 9a-9h. In no-go trials (Figs. 9g,9h), all neurons fire near their baseline rates, as prescribed (Equation 9). Thus, in this condition the network ignores the stimuli. In go trials, however, the profile of output responses has peaks at -10 and +10, which are the only two target locations in this task. In contrast to the activity profiles seen in previous tasks, here there never is a unique peak, even with zero noise (Figs. 9a,9c,9e). Instead, the relative amplitude of the two peaks varies as a function of bar orientation. The difference in the amplitudes of the two hills of activity decreases as the bar becomes more vertical, thus reflecting the difficulty of the task. Without any noise, the largest peak is always located at the correct target, but with noise the amplitudes vary across trials and errors are produced (Fig. 9d).

To quantify the performance of the network in this case, the generated movement was set equal to the location of the tallest hill of activity. This always corresponded to one or the other target location, +10 or -10, so each trial could be scored as either correct or incorrect. The assumption here is that a profile of activity with two peaks, as in Figs. 9c,9d, can be converted into a profile with a single peak, such that the smaller hill of activity is erased. Networks with recurrent connections organized in a center-surround fashion can do just that [26,34-36]. So, if such lateral interactions were added to the output layer of the network, only the largest hill of activity would remain. Equivalently, the responses of the output neurons could serve as inputs to an additional, third layer that performed the single-target selection [34]. Either way, given that this is a plausible operation, it is reasonable to simply consider the location of the largest peak to determine the evoked movement.

Based on this criterion, the performance of the network is shown in Figs. 9i,9j, which plot the probability or fraction of movements to the target on the right as a function of stimulus orientation $x$. These are essentially neurometric curves – psychometric curves computed from neuronal responses – and indeed have the sigmoidal shape that is characteristic of many psychophysical measurements. Figure 9i shows the results for condition 1, in which bars with $x > 0$ correspond to movements to the right; Fig. 9j shows the results for condition 2, in which the association is reversed and bars with $x > 0$ correspond to movements to the left. The gray curves are best fits to the simulation data points. The fits have two parameters, the center point, or bias (indicated by dashed lines), and a second parameter that determines the steepness at the center point and is inversely proportional to the discrimination threshold (see Methods). Without noise, performance is virtually perfect (not shown), in which case the bias and threshold are zero and the neurometric curve becomes a step function. However, both quantities increase in magnitude as noise is increased, producing the observed sigmoidal curves.

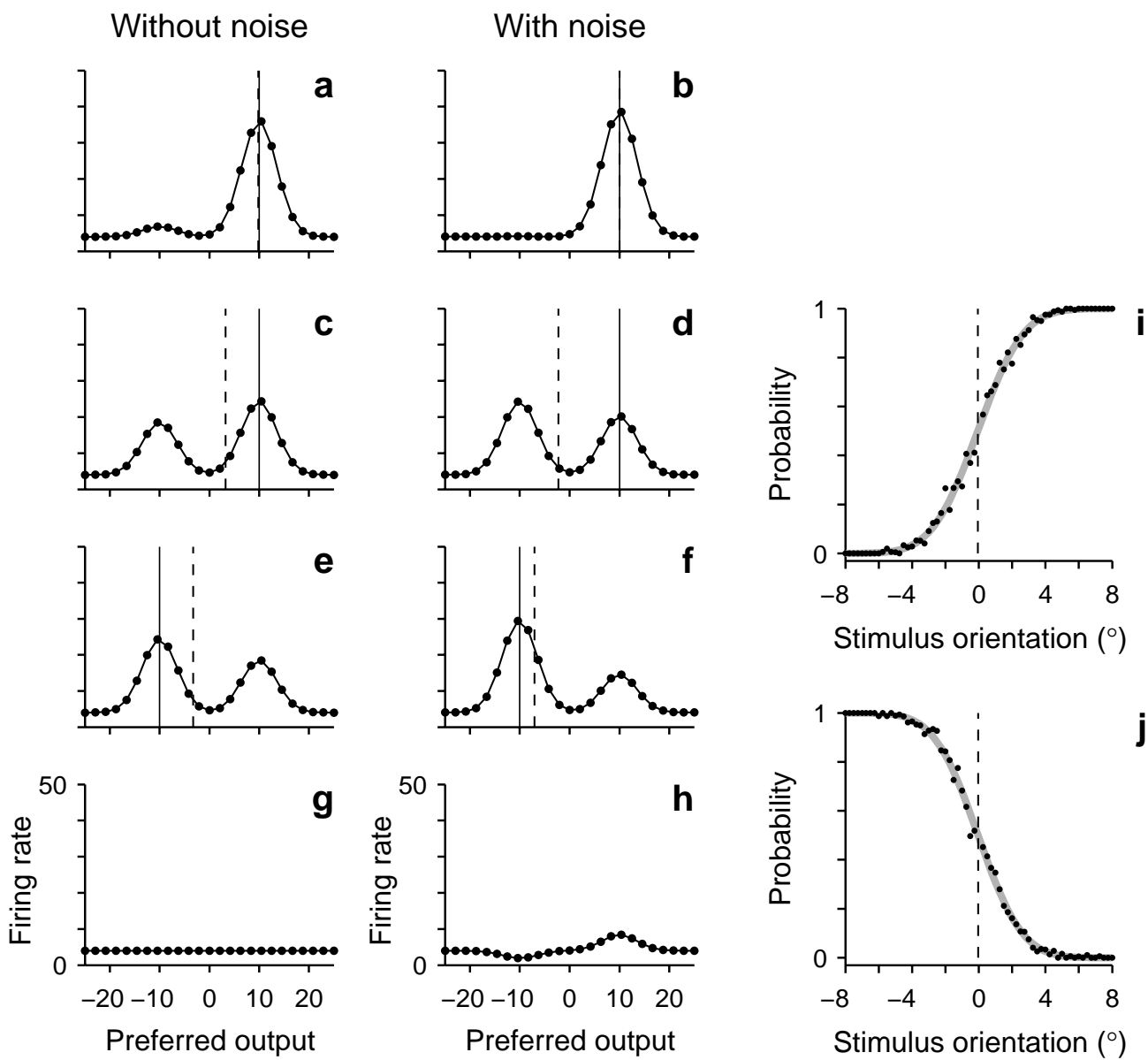

**Figure 9**
**Network performance during orientation discrimination.** Panels **a-h** show all 25 output responses, as driven by the GM neurons (not shown), in single trials. Continuous lines indicate intended target location (-10 or +10); dashed lines indicate the center of mass of the output activity. Right and left columns have identical stimuli and conditions, but with ($\alpha$ = 1) and without ($\alpha$ = 0) noise, respectively. A trial is deemed correct if the higher peak of activity is situated at the intended target location. **a**: Single trial with $x$ = 5°, $y$ = 1 and error = 0.15. **b**: As in **a**, but error = 0.001. **c**: Single trial with $x$ = 1°, $y$ = 1 and error= 6.7. **d**: As in **c**, but error = 12.2. The response is scored as incorrect because the tallest hill of activity is not at the intended target. **e**: Single trial with $x$ = 1°, $y$ = 2 and error = 6.7. **f**: As in **e**, but error = -2.9. **g, h**: No-go trials. **i**: Probability of making a movement to the right target as a function of stimulus orientation, in condition 1 ($y$ = 1) and with noise. Gray lines are fits to the simulation data. The center point or bias of the fit is indicated by the dashed line and is equal to -0.06°. Discrimination threshold is 1.5°. **j**: As in **i**, but with the association between orientation and targets reversed ($y$ = 2). The center point is -0.04°; the discrimination threshold is 1.4°.

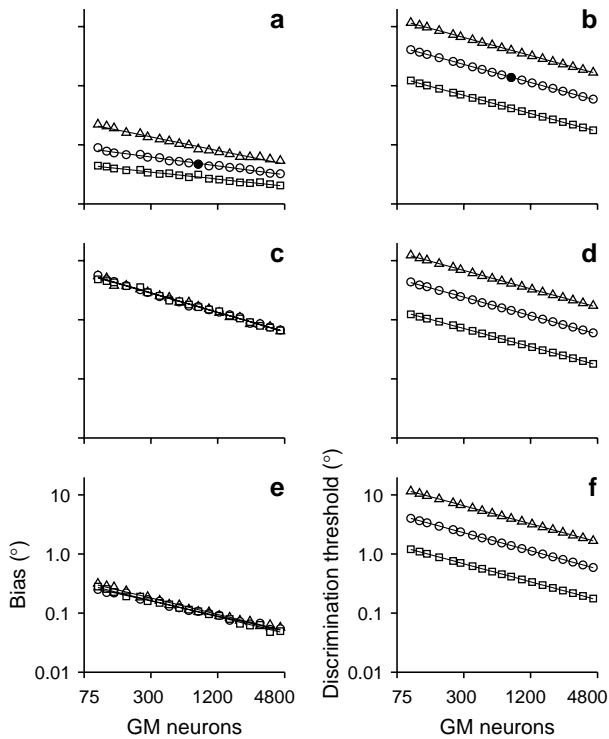

**Figure 10**
**Robustness and generalization in the orientation discrimination task.** Left and right columns show the bias and discrimination threshold, respectively of the neurometric fits (as in Figs. 9i,j) as functions of network size. **a, b**: Bias and discrimination threshold under standard conditions, which include 64 orientations used to set the connections and test the model. For each network size, results are absolute values averaged over the two go conditions and multiple networks. Filled symbols indicate the network used in Fig. 9. **c, d**: As in **a, b**, except that performance was tested after deleting 25% of the synaptic connections, chosen randomly. **e, f**: As in **a, b**, but when only 2 stimulus orientations (-8° and +8°) are used to set the connections, in combination with the 3 possible contexts, and performance is tested with all 64 orientations and 3 contexts. Straight lines are fits to the data points above 250 units.

The presence of a bias might be surprising, given the symmetry of the network. However, the bias depends on the number of trials used to estimate the probabilities. If each orientation were tested an infinite number of times, the data points in Figs. 9i,9j would line up perfectly along continuous curves. The discrimination thresholds of those curves would not be significantly different from those shown, but their biases would be zero. With finite samples, a bias in the neurometric curve is inevitable.

Figures 10a,10b show the bias and discrimination threshold as functions of network size for three levels of noise. Both quantities decrease with network size, so in this sense, the network is just as effective as that for the scaling task. Because large numbers of trials were used, the bias is about an order of magnitude smaller than the threshold. Figures 10c,10d plot the results when the synaptic connections in the network are corrupted by deleting 25% of them at random, as in Figs. 7c,7d. This manipulation leaves the discrimination threshold virtually unchanged, but increases the bias by about an order of magnitude, making it comparable to the threshold. This bias is a true limitation of the network; it does not decrease with more trials. Figures 10e,10f show performance during generalization, as in Fig. 7e. In this case, only the two extreme orientations, -8° and +8°, were used to set the connections (in combination with the three possible conditions). The network was then tested on the standard set of 64 orientations. A true bias also appears in this case. It stays lower than the threshold, which remains essentially unchanged.

In summary, although the ambiguity of the sensory information is reflected in the motor responses, it does not interfere with the context-dependent selection mechanism.

## Discussion
### Gain modulation as a switch
The above results demonstrate that contextual modulation could serve to select one of many associations or maps between sensory stimuli and motor responses. Indeed, a key insight is that modulating the gain of a neural population is, in a sense, equivalent to flipping a switch that turns on or off specific subpopulations of neurons. This explains why networks of GM neurons can generate large changes in downstream responses – even all-or-none changes, as in go vs no-go conditions (Figs. 9a–9h) – although their own activity may vary rather subtly. In this framework there is a distinction between the selection process and the sensory representations. The capacity to switch depends on the collection of gain factors, whereas the space of possible functions of the stimulus that can be computed downstream is determined primarily by the sensory tuning curves (Appendix A). A weaker modulation typically increases the sensitivity to noise of the resulting motor responses (Appendix C), but otherwise, partial modulation can achieve the same sensory-motor map selection as maximal, all-or-none modulation. This is why the mechanism works across a large variety of tasks and representations that involve some type of switch.

In a landmark paper, Pouget and Sejnowski [13] studied the capacity of GM networks for coordinate transformations using the concept of basis functions. A group of functions of $x$ form a basis set when any arbitrary function of $x$ can be computed as a linear superposition of those functions in the group; sines and cosines of are a well known example. The function of $x$ typically associated with a neuron's response is its tuning curve – its firing rate measured as a function of $x$. Pouget and Sejnowski showed that, starting with two networks that form separate basis sets for $x$ and $y$, a network of GM neurons comprising all possible combinations (i.e., pairwise products) of those two sets would form a basis set for functions that depend simultaneously on $x$ and $y$. This means that any function of $x$ and $y$ can be computed from the resulting GM responses. This was a crucial result, because it provided a rationale for generating such a combined representation. However, it assumed that both the sensory and modulatory variables are continuous and that, taken independently, the sets of $x$- and $y$-dependent tuning curves both form true basis sets.

The present results relax some of these assumptions and provide a complementary point of view. When the modulatory quantity $y$ varies discretely, each of its values corresponds to computing a different function of the stimulus $x$. Furthermore, the $x$- dependent tuning curves determine what functions of $x$ or sensory-motor maps can be computed downstream, but there is no requirement for them to form a strict basis set. As mentioned, the discontinuous case fits better with the idea of switching between various possible maps, as if separate populations of neurons were turned on and off. This approach also highlights two important characteristics of these networks, that the modulation factors need to be nonlinear functions of context (Appendix A), and that the sensitivity to noise depends on the magnitude of the modulation (Appendix C).

### Relation to other models

An important property of networks of GM neurons is that the output units read out the correct maps using a simple procedure, a weighted sum [13-15]. As a consequence, the overall strategy of these networks can be described as follows: the input data are first projected onto a high-dimensional space, and the responses in this space are then combined through much simpler downstream units that compute the final result – in the present case, $x$ and $y$ are the inputs and the high-dimensional space is composed of the GM responses. Interestingly, such expansion into an appropriate set of basis functions [13,37] is the central idea of many other, apparently unrelated models. For instance, this scheme is a powerful technique for tackling difficult classification and regression problems using connectionist models [33]. It also works for calculating non-trivial functions of time using spiking neurons [38]. This strategy might constitute a general principle for neural computation [37]. In addition, these networks are capable of generalizing to new stimuli and are quite resistant to changes in the connectivity matrix, so they don't require exceedingly precise fine-tuning.

### The problem of high dimensionality

A crucial requirement for the above scheme of projecting the data onto a suitable set of basis responses is to cover all relevant combinations of sensory stimuli and modulatory signals in the GM array [13-15]. It is the potentially large number of such stimulus-context combinations that may pose a challenge for these networks, a problem sometimes referred to as the curse of dimensionality [33]. In terms of the antisaccade task, for example, the context could be signaled by the shape or color of the fixation spot, the background illumination of the screen, a sound, or simply by past events, as would happen if the experiment ran in blocks of saccade and antisaccade trials. Each one of these potential cues would need to have a similar modulatory effect on the sensory responses, and it is not clear how the brain could establish all the necessary connections for this. Part of the problem is that we don't know how many independent dimensions there are – independence being the crucial property. For instance, the model for the antisaccade task has two contexts and requires two populations of switching neurons. More neurons are needed to deal with the version of the task that has five scales or contexts, but the number of necessary neurons does not keep growing endlessly; if the modulatory terms are chosen appropriately, a relatively small number of neurons can generalize to any scale, in effect generating an infinite number of sensory-motor maps. Of course, the key is that these are not independent, so the network can generalize. Thus, the scheme might work with realistic numbers of neurons if the number of independent context dimensions is not exceedingly large, but estimating this number is challenging.

**Table 1: Errors in performance for various possible interactions between stimulus and context.**

| Error/Task | *fg* | *f* + *g* | [*f* + *g* - 1]$_+$ | sig (*f* + *g*) | $a_p(f+g)^{b_p}$ |
|---|---|---|---|---|---|
| ($\sigma_{CM}$/Scaling (DE) | 0.60 | 6.3 | 0.50 | 0.62 | 0.66 |
| ($\sigma_{CM}$/Scaling (CE) | 0.60 | 5.5 | 0.51 | 0.61 | 0.69 |
| Bias/Orientation | 0.03° | 36° | 0.05° | 0.04° | 0.06° |
| DT/Orientation | 1.39° | 37° | 1.09° | 1.44° | 1.88° |

Functions *f* and *g* are the sensory- and context-dependent terms used to generate the GM responses; *fg* is the standard condition in which these functions are multiplied (Equation 1). Other combinations: *f* + *g*, linear interaction (Equation 15); [*f* + *g* - 1]$_+$, rectification (Equation 16); sig(*f* + *g*), sigmoidal function (Equation 17); and $a_p(f+g)^{b_p}$, power function (Equation 18). Network parameters were as in Figs. 6 and 9a-f, for the corresponding tasks. For each row, all numbers were generated using exactly the same model parameters, except for the specific combination of *f* and *g* terms. DE, discontinuous encoding; CE, continuous encoding; DT, discrimination threshold.

Another possibility is to have a relatively small number of available gain modulation patterns controlled by an additional preprocessing mechanism that would link them to the current relevant cue (the color of the fixation spot, its shape, the background illumination, etc.), a sort of intermediate switchboard between possible contexts and possible gain changes. Attention has some features that fit this description – it can select or favor one stimulus over another, it can act across modalities, and it can produce changes in gain [21-23,39]. No specific proposals in this direction have been outlined yet, neither theoretically nor experimentally, but this speculative idea deserves further investigation.

*Is exact multiplication needed?*
A key ingredient of the general, two-layer model is that the GM neurons must combine sensory and modulatory dependencies, *f*(*x*) and *g*(*y*), nonlinearly [13-15]. Results of two manipulations elaborate on this. First, when *f* and *g* were added (Equation 15) instead of multiplied, all transformations failed completely, as expected [13]. Second, when the sensory- and context-dependent terms were combined using other nonlinear functions (a sigmoid function, a rectification operation or a power-law; see Equations 16–18), accuracy remained approximately the same in all tasks. Results are shown in Table 1, which compares the performance of networks that implemented different types of stimulus-context interactions but were otherwise identical. This shows that the exact form of the nonlinearity used to combine *f* and *g* is not crucial for these models.

However, in some cases a multiplication allows the synaptic connections to be learned through simple Hebbian mechanisms [14,15], so it may be advantageous for learning. At least under some conditions, neurons combine their inputs in a way that is very nearly multiplicative [11,21-23]. Perhaps they do so when multiplication provides a specific computational advantage.

*Mixed sensory-motor activity*
In the model for the orientation discrimination task, the level of activity of the output neurons reflects not only the evoked movement but also the difficulty of the sensory process. This is consistent with the observation that, during sensory discrimination tasks, neuronal responses in many motor areas carry information about the stimulus [40-42]. This activity is often interpreted as related to a decision-making process. In the discrimination model, the responses of the neurons encoding the movement toward one of the targets increased in proportion to the strength of the sensory signal linked with that target (Figs. 9a-9f), as observed experimentally [40-42]. The model was not designed to do this. It simply could not generate single, separate peaks of activity for two nearby orientations on the basis of a single feedforward step; an additional layer or additional lateral connections would be required for that. Nevertheless, when such selection mechanism is assumed to operate, remapping proceeds accurately, even when the strength of the sensory signal varies. According to the model, sensory and motor information should be expected to be mixed together when distinct, non-overlapping responses (e.g., movement to the left or to the right) are generated on the basis of small changes in a stimulus feature that varies continuously, as orientation did in this task.

*Responses that depend on multiple cues*
In the present framework, if sensory responses were modulated by multiple environmental cues, the responses of downstream neurons could be made conditional on highly specific contextual situations (see ref. 16). Therefore, this mechanism may also explain the capability of some neurons to drastically change their response properties in a context-dependent way. Two prominent examples are hippocampal place cells, whose place fields can be fully reconfigured depending on multiple cues [43,44], and parietal visual neurons, which become selective for color only when behavioral context dictates that color is relevant [45].

In many tasks, two or more inputs are combined into conditional statements – 'if *X* and *Y* then *Z*'. The switching property of GM networks is useful in these situations as well. The study of abstract rule representation by Wallis and Miller [29] is a good example. In their paradigm, the decision to hold or release a lever depends on an initial cue and on two pictures. The cue indicates which of two rules, 'same' or 'different', is applied to the pictures. If the rule is 'same', the lever is released when the two pictures are identical but not when they are different; if the rule is 'different' the situation reverses, the lever is released when the two pictures are different but not when they are identical. To execute the proper motor action, two conditions must be checked. With the framework presented here, it is straightforward to build a model for that task; all it requires is a neural population that encodes the similarity of the pictures (i.e., is selective for matching vs non-matching pairs) and is gain modulated by the rule. Although the exact form of the modulation, for instance, whether it is close to multiplicative, is hard to infer from their data, the findings of Wallis and Miller [29] are generally consistent with the types of responses predicted by the model.

### *Experimental predictions*

Other experimental studies also include results that are consistent with gain interactions between multiple sensory cues [2,27,28,30] or with gain changes due to expected reward [46]. Interpreting these data is problematic, however, because those experiments were not designed to test whether changes in context generate changes in gain. The tasks described here, or similar paradigms, may be simplified to eight or so stimuli and two or three conditions, generating stimulus sets that would be within the range of current neurophysiological techniques with awake monkeys. The key is to be able to construct full response curves (Figs. 2,5), so that neuronal activity across contexts can be compared for several stimuli – not only for two, as is often done. This is because the models make three basic predictions that can only be tested with multiple stimuli and conditions: the responses should have mixed dependencies on stimulus and context, the mixing should be nonlinear, and the neurons should behave approximately as a basis-function set, in the sense that a weighted sum of their responses should approximate an arbitrary function of stimulus and context extremely well [31,47].

Ideally, the nonlinear mixture will show up through multiplicative changes in gain, as in Figs. 2 and 5, where the context-dependent variations in firing intensity respect stimulus selectivity. This could certainly happen [11,21-23], especially for some individual neurons, but other nonlinearities are possible [19,47,48] and might work equally well. A key observation is that context can include widely different types of circumstancial information, such as expected reward, motivation, fear or social environment [1,5-7]. Therefore, given the versatility of the models discussed here, a broader implication of the present work is the possibility that, as a basis for adaptive behavior, the brain systematically creates sensory responses that are nonlinearly mixed with numerous types of contextual signals.

## Conclusions

The framework discussed here demonstrates how to make a neural network adaptable to various environmental contingencies, labeled here simply as context. To achieve this flexibility, context must influence the ongoing sensory activity in a nonlinear way. This strategy was illustrated with tasks akin to those used in neurophysiological experiments with awake monkeys, but is generally applicable to the problem of executing a sensory-evoked action only when a specific set of conditions are satisfied. The mechanism works because changing the gain of multiple neurons is, in a sense, equivalent to flipping a switch that turns on and off different groups of neurons. Its main disadvantage is that all relevant combinations of stimulus and context must be covered, which may require a large number of units. On the upside, however, the switching functionality is insensitive to the quality or content of the sensory signals, is robust to changes in connectivity, and places minimal restrictions on how context is encoded. Future experiments should better characterize how cortical neurons integrate sensory and contextual information.

## Methods

### *Neuronal responses*

The GM responses depend on a sensory feature *x*, which may represent stimulus location or stimulus orientation, and on a context signal *y*. For each model GM cell, these quantities are combined through a product of two factors, *f* and *g*. The former determines the sensory tuning curve of the neuron and the latter its gain or amplitude as a function of context *y*. The mean firing rate $r_j$ of GM unit *j* is thus written as

$$r_j = r_{max} f_j(x) \, g_j(y) + B, \quad (1)$$

where $f_j$ and $g_j$ vary between 0 and 1, *B* is a baseline firing rate equal to 4 spikes/s and $r_{max}$ = 35 spikes/s. The specific functions used for $f_j$ and $g_j$ depend on the task and are described below. However, note that because these two terms are combined through a multiplication, changes in context alter the overall responsiveness of a cell, but not its selectivity, which is the defining feature of gain modulation [8,9]. To include neuronal variability, Gaussian noise is added to all GM responses in each trial of a task. The noise is multiplicative; that is, the variance of the noise for unit $j$ is equal to $\alpha r_j$, where $\alpha$ is a constant. Qualitatively similar results are obtained with additive instead of multiplicative noise.

The output neurons form a population code that represents the location of an impending eye movement. Each firing rate is determined by a weighted sum of the GM rates, where the weights correspond to synaptic connections. The mean rate of output neuron $i$ is

$$R_i = \sum_j w_{ij} r_j, \qquad (2)$$

where $w_{ij}$ is the synaptic weight from GM neuron $j$ to output neuron $i$. The output rates should encode the location of the movement to be made in each trial, so their profile of activity should have a single peak indicating the location that should be reached. The synaptic connections that achieve the correct sensory-motor alignment are found through an optimal algorithm, which is described further below.

Equation 2 is used when the GM neurons drive the output neurons. But for each task, there is also an intended or desired response for each output unit. This is denoted as $F_i$, and is a function of the stimulus and the context. Thus, $R_i$ and $F_i$ refer to the same postsynaptic neuron, but one is the actual driven response and the other is the intended response. The functions $f$, $g$ and $F$ vary across tasks, as described next.

### Parameters for the antisaccade task

In this task (Fig. 1), stimuli appear at a location $x$ in two possible contexts, labeled $\gamma = 1$ and $\gamma = -1$. The response should be a movement toward the location equal to $x\gamma$. The firing rates depend on the following functions. The tuning curves are Gaussians,

$$f_j(x) = \exp\left(-\frac{(x-a_j)^2}{2\sigma_f^2}\right), \qquad (3)$$

with the preferred stimulus location $a_j$ between -25 and +25 and $\sigma_f = 4$ (in Figs. 2,3) or $\sigma_f = 6$ (everywhere else). Because there are two conditions, the modulatory functions $g_j$ take only 2 values, 1 and $\gamma$, where $\gamma$ is the minimum gain. Crucially, one half of the GM neurons have $g_j(\gamma = 1) = 1$ and $g_j(\gamma = -1) = \gamma$, whereas the other half have the opposite context preference, $g_j(\gamma = 1) = \gamma$ and $g_j(\gamma = -1)$ = 1. The only exceptions are Figs. 3c,3f, in which the gain factors $g_j$ for each neuron were chosen randomly from uniform distributions. In this task, the desired response of output neuron $i$ is

$$F_i(x,\gamma) = r_{max} \exp\left(-\frac{(x\gamma - c_i)^2}{2\sigma_F^2}\right) + B, \qquad (4)$$

where $c_i$ is the preferred movement location of unit $i$ and $\sigma_F = 4$. Therefore, the output profile of activity (obtained by plotting $F_i$ vs $c_i$) should be a Gaussian centered at the intended target location, $x$ or $-x$, depending on the context.

### Parameters for the scaling task

The scaling task is identical to the antisaccade task, except that the context $\gamma$ can take many values. The tuning functions $f_j$, are the same (Equation 3) and the output responses again encode the location given by $x\gamma$ (Equation 4). The gain factors depend on which of two possible representations is used. When context is encoded discontinuously, five scales are used, $\gamma$ = -1, -0.5, 0, 0.5 or 1, so the modulatory functions $g_j$ must take five values; these are

$$g_j(\gamma) = \{1, 0.9, 0.75, 0.65, 0.5\}. \qquad (5)$$

Crucially, they are assigned randomly to each of the 5 conditions, with a new random permutation for each GM unit. As a final step, the $g_j$ values are jittered by small, random amounts (see Figs 5a,5b). On the other hand, when context is encoded continuously, each neuron is assigned a preferred scale $b_j$ between -1.4 and +1.4, and its gain is a Gaussian function of $\gamma$,

$$g_j(x) = \frac{1}{2} + \frac{1}{2}\exp\left(-\frac{(\gamma - b_j)^2}{2\sigma_g^2}\right), \qquad (6)$$

with $\sigma_g = 0.3$. Note that the minimum gain is 0.5 in both cases.

### Parameters for the orientation discrimination task

In this task (Fig. 8), $x$ is the orientation of a bar and varies between -8° and +8°, where $x = 0$° corresponds to vertical. The discrimination can occur in two ways: either left and right targets correspond to left- and right-tilted bars, respectively ($\gamma$ = 1), or viceversa ($\gamma$ = 2). In addition, there is a no-go condition ($\gamma$ = 3), for a total of three contexts. The orientation tuning curves are given by cosine functions,

$$f_j(x) = \frac{1}{2}(1 + \cos(2(x-a_j))) , \qquad (7)$$

where $a_j$ is now a preferred orientation between -90° and +90°. The modulation functions $g_j$ are generated as in the discontinuous version of the scaling task, except with three values,

$$g_j(y) = \{1, 0.75, 0.5\}. \quad (8)$$

The order in which each GM neuron prefers the three contexts is random. The responses of the motor neurons are given by

$$F_i(x,y) = \begin{cases} r_{max} \exp\left(-\frac{(M(x,y)-c_i)^2}{2\sigma_F^2}\right) + B & \text{if } y \neq 3 \\ B & \text{if } y = 3 \end{cases} \quad (9)$$

with $y = 3$ being the no-go condition and $\sigma_F = 4$. Here, $M(x, y)$ is the correct movement location, either -10 or +10, when orientation $x$ is presented in condition $y$. Specifically, for $y = 1$, $M = -10$ if $x < 0$ and $M = +10$ if $x > 0$; and for $y = 2$, $M = -10$ if $x > 0$ and $M = +10$ if $x < 0$. In no-go trials, all output responses should stay at the baseline level, $B$.

Simulation results in the orientation discrimination task are presented in terms of the probability of generating a movement toward the target on the right, $P_R(x)$, which is a function of orientation. Those results are fitted to the curve

$$P_R(x) = \frac{1}{2}\left(1 + \text{erf}\left(\frac{(x-a_e)}{b_e}\right)\right), \quad (10)$$

where erf is the error function. This expression has two parameters: $a_e$, which is the center point, or bias, and $b_e$, which is inversely proportional to the maximum slope. The discrimination threshold is defined as one half of the difference between the values of $x$ that give $P_R = 0.75$ and $P_R = 0.25$; for Equation 10 it is equal to $b_e$ erfinv(1/2), where erfinv is the inverse of the error function.

### Calculation of synaptic weights

The synaptic weights are chosen so that, on average, the driven and desired responses of the output neurons are as close as possible. This means that

$$\left\langle \sum_i \left(\sum_j w_{ij} r_j - F_i(x,y)\right)^2 \right\rangle \quad (11)$$

must be minimized. As in Equation 2, $w_{ij}$ is the connection from GM neuron $j$ to output neuron $i$. The angle brackets indicate an average over all values of $x$ and $y$ and over multiple trials. The optimal connections are found by taking the derivative of the above expression with respect to $w_{pq}$, setting the result equal to zero, and solving for the connections. The result is

$$w_{ij} = \sum_k L_{ik} C_{kj}^{-1}, \quad (12)$$

where

$$C_{kj} \equiv \langle r_k r_j \rangle \quad (13)$$

$$L_{kj} \equiv \langle F_k(x,y) r_j \rangle. \quad (14)$$

Equations 12–14 are the recipe for setting the connections. Here **C**$^{-1}$ is the inverse of the correlation matrix **C** defined above. This inverse (or the pseudo-inverse) is found numerically. To calculate the averages defined above, the GM rates for all values of $x$ and $y$ are needed. These are found by evaluating Equation 1 plus the noise term for each GM neuron. When the noise is uncorrelated across neurons, as in the simulations, it only contributes to the diagonal of **C**. Because the variance of the noise is equal to $\alpha$ times the mean response, noise adds an amount $\alpha \langle r_i \rangle$ to element $C_{ii}$ of the correlation matrix. Except for this, all averages $C_{kj}$ and $L_{kj}$ are obtained from the mean input responses given by Equation 1 and the corresponding $F_i$ functions of the output neurons.

Having specified the parameters of the network (number of GM and output units, tuning and gain functions, stimulus-movement association), the procedure for setting the synaptic weights is run only once. Afterward, the connections are not adjusted any further.

### Other response functions

Equation 1 is based on an exact multiplication between $f_j(x)$ and $g_j(y)$. The effects of other possible interactions between stimulus and context are investigated using four alternative expressions in place of Equation 1. First, a linear combination of sensory and context signals,

$$r_j = \frac{r_{max}}{2}(f_i(x) + g_j(y)) + B. \quad (15)$$

Then, three nonlinear interactions. The first one is based on rectification,

$$r_j = r_{max} [f_j(x) + g_j(y) - 1]_+ + B, \quad (16)$$

where $[x]_+ = \max\{0, x\}$. The second one uses a sigmoid function,

$$r_j = \frac{r_{max}}{1 + \exp(-(f_j(x) + g_j(y) - a_s)/b_s)} + B, \quad (17)$$

The sigmoid is widely used in artificial neural networks [12] and has two parameters, $a_s$ and $b_s$. The third nonlinear interaction is based on a power law,

$$r_j = r_{max} a_p (f_j(x) + g_j(\gamma))^{b_p} + B, \quad (18)$$

and has two parameters too. This type of expression approximates some of the gain effects observed experimentally [49]. The free parameters in these expressions are adjusted so that the resulting firing rates are as close as possible to those given by Equation 1. All else is as in the original simulations.

## Outline of the simulations

Having specified a task, the tuning and gain curves of the GM neurons, and the network connections, the model is tested in a series of trials of the task. Each trial consists of the following steps: (1) specifying the stimulus and context, $x$ and $\gamma$, (2) generating all GM responses (Equation 1), (3) calculating the driven, output responses (Equation 2), and (4) determining the encoded movement $M_{out}$ by using the center of mass of the motor activity profile (Equation 19). Finally, the encoded movement is compared to the movement $M_{desired}$ that should have been performed given $x$ and $\gamma$ – their difference is the error in that particular trial.

The encoded movement $M_{out}$ is equated with the center of mass of the output population,

$$M_{out} = \frac{\sum_i (R_i - B)^2 c_i}{\sum_k (R_k - B)^2}, \quad (19)$$

where $c_i$ is the preferred target location of output unit $i$. The root-mean-square average of the motor error is used to quantify performance over multiple trials,

$$\sigma_{CM} = \sqrt{\langle (M_{out} - M_{desired})^2 \rangle}, \quad (20)$$

where only go trials are included in the calculation. On average, the encoded movement is very near the desired one, $\langle M_{out} - M_{desired} \rangle \approx 0$. Thus, $\sigma_{CM}$ is the standard deviation of the motor error, and measures the accuracy of the output population as a whole.

In all tasks, 25 output units are used, with $c_i$ uniformly spaced between -25 and 25. Preferred stimulus values $a_j$ and preferred context values $b_j$ are first distributed uniformly and then jittered by small, random amounts.

All simulations were performed using Matlab (The Mathworks, Natick, MA). The source code is available on request.

## Appendix A

This section shows that, with a finite number of contexts, gain modulation is functionally equivalent to a switch. More precisely, for a discrete number of contexts and everything else being equal, a network of partially modulated neurons can generate the same mean downstream responses as a network of switching neurons.

Consider $M$ populations or groups of sensory neurons with identical sets of tuning functions $f_j(x)$. There are $N$ neurons in each population, so index $j$ runs from 1 to $N$. These populations project to a postsynaptic neuron through synaptic connections $w_j^p$, where the superscript indicates the presynaptic population of origin. Thus, $w_j^p$ is the synaptic weight from neuron $j$ in group $p$ to the postsynaptic unit. The sensory neurons are gain modulated, so the mean response of unit $j$ in population $p$ is given by

$$r_j^p(x, \gamma) = g_j^p(\gamma) f_j(x), \quad (21)$$

where $x$ and $\gamma$ label the stimulus and the context, as before. Next, assume that there are $M$ possible contexts, so $\gamma$ can take integer values from 1 to $M$. Therefore, the gain factors can be expressed as three-dimensional arrays, and the presynaptic firing rates can be rewritten as

$$r_j^p(x, \gamma = k) = g_j^{pk} f_j(x). \quad (22)$$

Here, $g_j^{pk}$ corresponds to the gain of unit $j$ in population $p$ during context $k$. With this notation, the response of the downstream neuron during context $k$ becomes

$$R(x, \gamma = k) = \sum_{p,j} w_j^p g_j^{pk} f_j(x), \quad (23)$$

where the sums are over all populations and all units in each population. Note that, for each index $j$, the coefficient in front of the tuning function is given by the product of an $M$-dimensional vector of weights times an $M \times M$ matrix of gain factors.

The essential idea is to compare the response of the postsynaptic unit under two conditions: when only one input population is active in any particular context (and all others are fully suppressed), and when the populations are only partially suppressed, with different combinations of gain factors for each context. For this, the hat symbol ^ is used to label all quantities obtained in the former case, with switching neurons; that is, the hat means 'obtained with full modulation'.

Full modulation occurs when the matrix of gain factors for all the units with index $j$ is equal to the identity matrix,

$$\hat{g}_j^{pk} = \delta_{pk} = \begin{cases} 1 & \text{if } p = k \\ 0 & \text{if } p \neq k \end{cases}. \quad (24)$$

According to this expression, for population 1, the gain is 1 when $\gamma = 1$ and is 0 otherwise; for population 2, the gain is 1 when $\gamma = 2$ and is 0 otherwise, and so forth. Substituting this expression in equation 23 gives the firing rate of the downstream unit when driven by switching neurons,

$$\hat{R}(x, \gamma = k) = \sum_j \hat{w}_j^k f_j(x). \quad (25)$$

Again, the hat simply indicates that the quantity was obtained with maximally modulated input neurons. In this case, $\hat{R}(x, \gamma = k)$ implements a different function of $x$ for each context value, such that the function expressed in context 1 depends only on the weights from the first population of switching neurons, $\hat{w}_j^1$, the function expressed in context 2 depends only on the weights from the second population, $\hat{w}_j^2$, and so on. This is the situation depicted in Figs. 3a,3d.

On the other hand, the postsynaptic response driven by partially modulated units is simply as in Equation 23, where the absence of a hat means 'obtained with partial modulation'. Under what conditions is the output response driven by partially modulated neurons, $R(x, \gamma = k)$, equal to the response obtained with full modulation, $\hat{R}(x, \gamma = k)$? Compare the right hand sides of Equations 23 and 25; for them to be the same, the coefficients in front of the tuning functions must be equal; that is,

$$\sum_p w_j^p g_j^{pk} = \hat{w}_j^k. \quad (26)$$

This condition is satisfied if the weights with partial modulation are set equal to

$$w_j^p = \sum_q \hat{w}_j^q h_j^{qp}, \quad (27)$$

where $\mathbf{h}_j$ is the inverse of the matrix of gain factors $\mathbf{g}_j$; that is, $\sum_p h_j^{qp} g_j^{pk} = \delta_{qk}$. Therefore, the key constraint here is that the gain factors in the partial modulation case must have linearly independent values across contexts, so that the inverses exist; in other words, the matrices $\mathbf{g}_j$ must have full rank. An important consequence of this is that for $M > 2$, the gain of each neuron as a function of context ($g_j^p(\gamma)$ in Equation 21) must be nonlinear.

Equation 27 is the key result. It provides a recipe for going from a network of switching neurons to a network of partially modulated neurons (given equal numbers and types of tuning functions $f_j(x)$). For the recipe to apply, the gain factors in the latter must have the appropriate inverses, but otherwise they are arbitrary. Because each possible function that a network can generate corresponds to a different matrix of synaptic connections, this implies that all the possible functions of $x$ that the output can implement with fully switching neurons can be replicated with partial gain modulation.

This statement is exact when there is no noise; with noise it applies to the average downstream responses. Notice that this result is independent of the tuning functions $f_j(x)$. These determine the possible functions of $x$ that can be generated downstream – that is, the available sensory-motor maps – but have no effect on how these are switched or selected. Finally, the result is also valid if the postsynaptic response is equal, not simply to the weighted sum of GM responses, but to an arbitrary function of that sum.

**Appendix B**

To illustrate the result in Appendix A, consider a simple case with two populations and two contexts, as in Figs. 1, 2, 3, 4. With full modulation, the response of the downstream neuron is

$$\hat{R}(x, \gamma = 1) = \sum_j \hat{w}_j^1 f_j(x) \quad (28)$$

in context 1, and

$$\hat{R}(x, \gamma = 2) = \sum_j \hat{w}_j^2 f_j(x) \quad (29)$$

in context 2. This is simply Equation 25 for $M = 2$. Context turns one sensory population on and another off. Now, how can we obtain the same downstream responses, as functions of $x$, when the GM neurons are partially modulated? First, suppose that the modulation matrices are

$$\mathbf{g}_j = \begin{pmatrix} 1 & \gamma \\ \gamma & 1 \end{pmatrix}. \quad (30)$$

The gain factors can only take two values, 1 for the preferred context, and $1 > \gamma \geq 0$ for the non-preferred one; the full-modulation case is recovered when $\gamma = 0$. For simplicity, these factors are the same for all units in each population, so there is no variation across index $j$. This matrix was used in Fig. 3, with $\gamma = 0$ and $\gamma = 0.5$ for the left and middle columns, respectively, and in Figs. 4a,4b. Its inverse is

$$\mathbf{h}_j = \begin{pmatrix} 1 & -\gamma \\ -\gamma & 1 \end{pmatrix} \frac{1}{1-\gamma^2}. \qquad (31)$$

Next, substitute into the transformation rule found earlier, Equation 27; the result is

$$w_j^1 = \frac{\hat{w}_j^1 - \gamma \hat{w}_j^2}{1-\gamma^2} \qquad (32)$$

$$w_j^2 = \frac{\hat{w}_j^2 - \gamma \hat{w}_j^1}{1-\gamma^2}. \qquad (33)$$

With these synaptic weights, the downstream responses driven by the partially modulated neurons (Equation 23 with $p = 1, 2$ and the gain factors in Equation 30) become identical to the rates driven by the switching neurons (Equations 28,29). This is the linear transformation used in Figs. 4a,4b.

## Appendix C

Using partial instead of full modulation to switch between maps does come at a price: the variability of the postsynaptic response typically increases. This can be seen as follows.

If additive noise is included in the input firing rates, the response of neuron $j$ in population $p$ becomes

$$r_j^p(x, y = k) = g_j^{pk} f_j(x) + \epsilon_j^{pk}, \qquad (34)$$

where $\epsilon_j^{pk}$ is a random fluctuation for unit $j$ in population $p$ during context $k$. The variance across trials of this random variable is denoted as $\sigma_r^2$ and is the same for all GM neurons. The downstream neuron has the same mean response as before (Equation 23), but now it has a variance, which is equal to

$$\sigma_R^2 = \left\langle \left( \sum_{p,j} w_j^p \epsilon_j^{pk} \right)^2 \right\rangle$$

$$= \left\langle \sum_{p,j,q,i} w_j^p \epsilon_j^{pk} w_i^q \epsilon_i^{qk} \right\rangle$$

$$= \sigma_r^2 \sum_{p,j} \left( w_j^p \right)^2. \qquad (35)$$

Here, the angle brackets indicate an average over trials, which affects the noise terms only. To go from the second to the third line above, the key is to assume that the fluctuations are independent across neurons, such that $\left\langle \epsilon_j^{pk} \epsilon_i^{qk} \right\rangle = \delta_{ij} \delta_{qp} \sigma_r^2$.

The next step is to compare the variance of the postsynaptic unit when driven by the switching neurons and by the regular, partially modulated GM neurons. For simplicity, consider the same 2 × 2 case as in Appendix B, where the modulation is parameterized by $\gamma$. The variance $\hat{\sigma}_R^2$ of the postsynaptic response driven by switching neurons is exactly as in Equation 35, but with $\hat{w}_j^p$ and $p = 1,2$. This must be compared to the variance obtained with partial modulation for the same mean postsynaptic responses. The synaptic weights that achieve this are given by Equations 32 and 33; substituting those into Equation 35 gives

$$\sigma_R^2 = \frac{1}{(1-\gamma^2)^2}((1+\gamma^2)\hat{\sigma}_R^2 - a\gamma), \qquad (36)$$

This is the variance of the postsynaptic response driven by an array of partially modulated GM neurons as a function of the variance $\hat{\sigma}_R^2$ obtained when the response is driven by fully modulated, switching neurons. Here, $a$ depends on the weights $\hat{w}_j^p$, but is not a function of $\gamma$,

$$a = b \sum_j \hat{w}_j^1 \hat{w}_j^2, \qquad (37)$$

where $b$ is a constant. Note that $a$ is a measure of the overlap between the sets of connections from the two populations. The dependence of $\hat{\sigma}_R^2$ on the weights is such that $a \leq 2\hat{\sigma}_R^2$.

Equation 36 shows that, although the average postsynaptic response is the same function of $x$ and $y$ for all $\gamma$, its variability changes with $\gamma$. A similar result is obtained when the variance of the input firing rates is proportional to their mean. In that case,

$$r_j^p(x, y = k) = g_j^{pk} f_j(x) + \alpha \sqrt{g_j^{pk} f_j(x)} \, \epsilon_j^{pk}, \qquad (38)$$

with $\sigma_r^2 = 1$. A calculation analogous to the one just described leads to

$$\sigma_R^2 = \frac{1+\gamma}{(1-\gamma^2)^2}((1+\gamma^2)\sigma_R^2 - a\gamma) \,, \qquad (39)$$

where $a$ is the same as in Equation 37, except with a different proportionality constant. In this case, it is still true that $a \leq 2\hat{\sigma}_R^2$. This expression was used to generate the continuous lines in Fig. 4a. For this, $\hat{\sigma}_R^2$ was simply the variance in the postsynaptic firing rate found from the simulations with $\gamma = 0$, and $b$ was chosen to generate the best fit to the rest of the simulation data points.

Equations 36 and 39 do not always increase monotonically with $\gamma$. This depends on $a$, which is a measure of the similarity between the sensory-motor maps established in the two contexts. For instance, when the two maps are the same, $\hat{w}_j^1 = \hat{w}_j^2$ for all $j$, and $a$ attains its maximum value, $2\hat{\sigma}_R^2$. In that case, the variance with partially modulated neurons is always smaller than with switching neurons. This makes sense: if the maps in the two contexts are the same, it is always better to have the two populations active at the same time, as this reduces the noise. According to the analysis, when $a = 2\hat{\sigma}_R^2$ and $\gamma = 1$, Equation 36 gives $\sigma_R^2 = \hat{\sigma}_R^2/2$. The variance is divided by 2 because noise is additive and there are two active populations doing the very same thing. In contrast, when the two maps are different, their respective synaptic weights are also different, and $a$ is either positive but much smaller than $2\hat{\sigma}_R^2$, or negative. Then, $\sigma_R^2$ might have a minimum for some intermediate value of $\gamma$, or may increase monotonically, which is what happens in Figs. 3 and 4, with saccades vs antisaccades.

### List of abbreviations
GM, gain-modulated.

### Acknowledgements
I thank Terry Stanford and Nick Bentley for useful discussions, Sacha Nelson for suggesting the calculation in Appendix C, and two anonymous reviewers for their comments. Research was supported by NINDS grant NS044894.